\documentstyle[aps,preprint]{revtex}

\input epsf.sty

\begin{document}

\title{Nonequilibrium effects in superconducting necks of nanoscopic dimensions}
\author{H. Suderow$^{1,2}$, S. Vieira$^{2}$}
\address{$^{1}$ Instituto de Ciencia de Materiales de Madrid, Consejo
Superior de Investigaciones Cient\'ificas, 28049 Madrid
\\
$^{2}$ Laboratorio de Bajas Temperaturas, Departamento de Fisica de
la Materia\\ Condensada, Instituto de Ciencia de Materiales
Nicol\'as Cabrera, Facultad de%
\\
Ciencias, C-III, Universidad Aut\'onoma de Madrid, 28049
Madrid-Spain}
\date{\today}
\maketitle

\tightenlines
\widetext

\begin{abstract}
We have fabricated superconducting connecting necks of Pb with a
scanning tunneling microscope (STM) and studied their properties
under magnetic fields near the transition to the resistive state. A
striking phenomenology is found with two well defined conduction
regimes as a function of the magnetic field. We discuss the
possible origin of this behavior in terms of the interplay between
the field dependence of the quasiparticle charge imbalance length
$\Lambda_{Q*}$ and the length of the neck which is superconducting
under field.

\bigskip

\center{TO BE PUBLISHED IN PHYSICS LETTERS A}

\end{abstract}

\pacs{PACS numbers:
74.40.+k,74.50+r,74.80.Fp}


The study of very small superconducting systems is a topical
research area showing many interesting aspects, relevant for
possible applications in nanoelectronics\cite{Review}. Here we deal
with connecting necks of nanoscopic dimensions between two
superconducting electrodes. We expect that their properties are
important to design of superconducting circuits at the smallest
length scales, as they can help to understand the behavior of the
connecting structures that are needed to operate those circuits. We
build the necks by repeatedly indenting the tip of a scanning
tunneling microscope into a sample (both being simple metals, like
Pb), having the form of a constriction with a smallest cross
section making the contact between both banks. These structures are
too small to be fabricated using present nanolithographic
techniques \cite{Cuellos,ModelU,OtherNecks}. The mechanical,
electronic transport, and superconducting properties have been
studied in
Refs.\cite{Cuellos,ModelU,OtherNecks,Rodr94,Poza98,Sud00}, where it
was realized that the simultaneous measurement of the conductance
and the displacement of the tip with respect to the sample give
information about the form of the neck. Depending on the position
on the surface where the repeated indentation is done, necks of
different forms are obtained in-situ. For example, long structures,
comparable to a short "nanowire" attached to bulk electrodes
\cite{ModelU}.  The smallest cross section of the neck, the last
contact, can be decreased towards the formation of a single atom
point contact \cite{Sud00,Atoms}. Transport is ballistic, as long
as the mean free path is larger than the diameter of the contact,
i.e. of the smallest cross section of the constriction\cite{PC}.

One of the most interesting problems in small scale superconductors
is the knowledge about their properties near the limits of the
 presence of superconductivity. For instance, the behavior of small wires
near the critical temperature with an applied current has been a
subject of intense research in the last decades, motivated both
from the applied and fundamental point of views. As regards
connecting necks, the transition at zero field turns out to be
easily understandable using simple models\cite{Rodr94}. Here we
present results under magnetic fields, where we find a surprising
phenomenology with anomalies in the differential resistance whose
presence depends on the geometry. We propose that these anomalies
are due to the fact that superconductivity is confined to
nanoscopic length scales under magnetic fields. Indeed, in previous
works we could show that in necks formed of Pb, which is a type I
superconductor, above the bulk critical field $H_c$,
superconductivity is destroyed in the banks, but the neck remains
superconducting when its lateral dimensions are smaller than the
London penetration depth or the coherence
length\cite{Poza98,Sud00}.

We use a STM within a $^4$He cryostat with a superconducting magnet
and a conventional four wire I-V technique together with a Lock-In
amplifier to measure the conductance.  Considerable care has been
taken to achieve a good resolution and to shield the apparatus from
RF\ noise. The experimental procedure to obtain clean necks has
been largely discussed\cite{Cuellos,ModelU,Rodr94,Poza98,Sud00},
but we repeat here the main points. We cut a tip and a sample from
a high purity (99,99\%) slab of Pb with a clean blade, immediately
mount them on the STM and cool down in $^4$He exchange gas.
Subsequently, we indent repeatedly the tip into the sample while
recording the resistance $R$ of the neck as a function of the
elongation of the piezotube $z$. $R(z)$ is a staircase function
whose overall form is different for each indentation and can be
related to the topological form of the neck using the model of
Ref.\cite{Checks,ModelU}. These authors found that one can model
the whole constriction by a series of small slabs, each one of
different radius and thickness, arranged symmetrically with respect
to the center.  We assume that the volume of the connecting neck
remains constant.  Then, the changes during the elongation of the
neck are interpreted by a plastic deformation of the slab with the
narrowest cross section which breaks and forms a new slab with a
smaller cross section. The smallest cross section $A$ as a function
of the elongation of the neck is easily calculated\cite{ModelU}. If
the resistance is near the Sharvin limit
($R_S=\frac{h}{2e^2}(\frac{\lambda_F^2}{\pi A})$ with $\lambda_F$
the Fermi wavelength) we can relate the measured $R=R_S$ to $A$ and
each measured $R_S(z)$ curve to a given $A(z)$ function containing
the form of the neck.  In this work, we choose $R\approx 5\Omega$,
which corresponds to a contact with a diameter of about $d=6$nm.
This is sufficiently small so that the conduction is near the
ballistic regime ($\ell>d$) but still large enough so that the neck
transits to the resistive state when a reasonably small current is
applied (Fig.1).  Note that in Ref.\cite{Sud00} we study the
conduction properties of similar necks with three orders of
magnitude smaller currents (the contact corresponding in that case
to a single atom).

The Fig.1 shows two $dV/dI(V)$ curves obtained at zero field
corresponding to two different geometries.  At zero voltage, a
Josephson current flows and above the critical current $I_{c,0}$
(limited by the contact), $dV/dI$ shows anomalies located at
$V_n=\frac{2\Delta }n$ with $n=1,2,3,....$ related to multiple
Andreev reflection processes\cite{Andreev} (arrows in Fig.\ 1). The
$I(V)$ curve is shifted with respect to the resistive state ohmic
I-V curve by the so called ''excess current'' $I_{exc}$ (which is
of the order of $I_{c,0}$ \cite{Andreev}), and smoothly disappears
when the neck is heated by the strong current flow. Therefore,
$dV/dI$ increases slightly above $2\Delta$ and decreases at larger
voltages, when the local temperature of the neck rises above
$T_c$\cite{Rodr94,Tinkhamheat}.  To estimate the local temperature,
one can use the position in voltage of the $n=1$ peak which is
indeed at smaller voltages than expected from the position of the
$n=2$ peak ($2V_2>V_1$ see arrows in
Fig.1)\cite{Rodr94,Tinkhamheat}.  Necks with larger opening angle
(open points in Fig.1) are not heated in the current range that we
study ($V_n$ agrees in that case with the above expression, see
also \cite{Sud00}, we measure $\Delta=1.35meV$).

Above $H_c$(Fig.\ 2), bulk superconductivity is destroyed, but the
neck remains superconducting \cite{Poza98,Sud00}. For $V \leq
2\Delta$, the curves behave as the zero field curve, but with the
features due to multiple Andreev reflections smoothed, in agreement
with Ref.\cite{Sud00} where we discuss this in detail
\cite{NotaXi}. But the results for $V>2\Delta$ show a completely
unexpected and very anomalous behavior.

High peaks appear in the $dV/dI$ curve above $2H_c$ for $2\Delta <
V<6\Delta $(Fig.\ 2).  The peaks appear on top of the distorted
$dV/dI$ curve due to the above mentioned heating effects.  A small
voltage step ranging from $10\mu V$ to $100\mu V$, i.e. a small
part ($0.2 - 2$\%) of the whole voltage drop, is associated with
each peak. The number of peaks varies between one and four. Each
time we sweep V (from $-15$mV to $+15$mV) at a given magnetic field
we do not find the peaks at the same voltage and their number is
also different. The effect is independent on the direction of
current flow. Note that in Fig.2 we only show the relevant voltage
range and some characteristic curves for clarity. At larger
magnetic fields (Fig.\ 2 c.\ and 2 d.), the number of observed
jumps decreases to one and moves towards slightly smaller voltages.
No peaks appear between $0.15$ T and $0.2$ T in this neck.  If we
build necks with larger opening angle $\theta$, or stretch the neck
towards smaller contacts, we also find a smooth behavior, without
peaks. Therefore, the observed phenomenology cannot be associated
to the (ballistic) conduction through the contact, but rather to
the contribution of a part of the neck which transits to the
resistive state under current flow. Indeed, in the necks with the
smallest opening angle (around $\theta\approx 30^{\circ}$, see
Fig.1), it reasonable to expect that the conduction is
quasiballistic, having a small contribution from regions of the
neck outside the smallest contact. The peaks clearly indicate that
the transition to the resistive state is discontinuous, and this
can be associated to the nucleation of phase slip centers within
the neck.

The signature of phase slips are steps in the V-I characteristics.
In phase slip wires,
 each {\ step} in the
V-I curve gives a peak in $dV/dI$, and $dV/dI$ increases by a
constant amount after every step
\cite{Kadin78,Klapwijk77,Scokpol74}, that is proportional to the
length $\Lambda _{Q*}$ where charge imbalance is found around the
phase slip. $\Lambda _{Q*}$ is in general much larger than the
coherence length and is given by \cite{Strunk,Schmid75}
$\Lambda_{Q*} = \sqrt{\frac 13v_F\ell \frac 4\pi
\frac{k_BT}{\Delta}\tau_{E-P}}$ with $\ell $ being the mean free
path, $v_F$ the Fermi velocity, $\tau_{E-P}$ the electron-phonon
scattering time, and $\Delta$ the superconducting gap. In our case,
we observe a small step in the $V-I$ curve and its associated peak
in $dV/dI$, because the voltage drop associated to the transition
to the resistive state of a part of the neck is in any case small.
In order to obtain a more precise estimation, we would need to
consider how the voltage drops on the whole neck, which is outside
the scope of this paper. To gain further understanding, it is
instructive to get an idea of $\Lambda
_{Q*}$ in our situation. In order to explain the presence of
several anomalies in our curves, we expect that $\Lambda _{Q*}$
should be of the order of or smaller than the maximal length of the
necks (200nm). Nevertheless, simple estimates give a larger value
for the lowest limit for $\Lambda
_{Q*}$ in Pb $\Lambda
_{Q*}> 430$ nm (we take $\tau _{E-P}\approx 2.5$ $10^{-11}s$
\cite{Schmid75} and $\ell\approx 10 nm$, see e.g.\cite{Strunk} for
the complete T-H dependence of $\Lambda_{Q*}$). Note that this is
in any case much smaller than $\Lambda _{Q*}$ in other phase slip
wires (which can easily reach microns, see
Refs.\cite{Kadin78,Klapwijk77,Strunk,Aponte83,Chien99}) as $\tau
_{E-P}$ is smaller in Pb than in other metals, but still larger than the
length of our necks. Under field, $\Lambda _{Q*}$ is further
reduced due to the pair breaking effect of the magnetic field which
introduces a new time scale for relaxation so that $\tau_{E-P}$ is
substituted by $\sqrt{\frac{\tau_{E-P}}{2
(1/\tau_{PB}+1/\tau_{E-P})} }$ in the expression given above for
$\Lambda _{Q*}$ ($\tau _{PB}$ is the pair breaking time; see
\cite{Strunk}).  In our case, the strong current flow should lead
to an additional decrease of $\Lambda_{Q*}$\cite{Chien99}. For
instance, in the field range of Fig.2, a small pair breaking
parameter (of $\frac{\hbar}{\tau_{PB}}=0.2T_c$), compatible with
the reduction of the superconducting gap with field \cite{Sud00}
already gives a reduction of $\Lambda_{Q*}$ by a factor of $2$. So
that $\Lambda_{Q*}$ can reach values comparable to the overall
length of our structures.  Although this gives only an order of
magnitude estimate of the effect of the magnetic field, it becomes
clear that the nucleation of a small number of phase slips within
the neck is favored by the field.  The irreproducibility in the
number and position of the jumps is also compatible with the
appearance of phase slips, as the situation is expected to be
complex in a locally overheated environment.

At about $0.2$ T, (Fig.\ \ref{fig:Fig3}) we observe another high
peak which becomes most clearly visible at $0.25$ T (Fig.3b) and
then drops (Fig.3c). The corresponding anomaly in the V-I curve is
again very small, of at most $ 100 \mu V$ (at 2.5kG) and it has
the same dependence on the geometry as the data shown in Fig.2,
i.e.  it is not present in necks with smaller contact or larger
$\theta$ (see Fig.3c).  We therefore also associate this peak with
the transition to the resistive state of the neck. The difference
with the peaks shown in Fig.2 is that, at a given field, when we
sweep $V$ (from $-15mV$ to $+15mV$) we reproduce the behavior
without changes nor hystheresis.

In this high field regime, we should take into account that the
length $L_S$ of the superconducting part of the neck decreases
significantly \cite{Sud00}. Indeed, we can separate the neck into
small disks each one having a critical field given by
$H_{critical}=4\lambda \frac{H_c}r$\cite{Poza98}. Then, $L_S$
decreases with field as $L_S\sim (8\lambda H_c)/(H\tan \theta )$.
According to \cite{Strunk}, $\Lambda _{Q*}$ decreases more rapidly
with field $\sim 1/\sqrt{H}$ in a simple, one dimensional system
\cite{comment}.  Therefore, as shown in the inset of Fig.\ 3c, at
the highest fields, the length of the superconducting part of the
neck can fall below $\Lambda_Q*$. At the same time, charge
imbalance created by the N-S interface leads to a greater voltage
drop as compared to lower fields because it is created near the
part of the neck with smaller radius.  This could lead to the
anomaly shown in Fig.3.  The recent calculations of
Ref.\cite{Sud00} on a similar neck confirm these estimations. It is
shown that at $0.25 T$ (the same field where the anomaly is at its
maximum), the N-S interface is well defined and located at a
distance about three times the coherence length from the center of
the neck.  At larger fields, the height of the corresponding
anomaly decreases as superconductivity is destroyed.

The whole phenomenology (Figs.\ 2 and 3) has been reproduced in
different experiments, with the magnetic field parallel or
perpendicular to the neck. It is robust and depends only on the
geometry of the neck.  We have measured several geometries (
$30^{\circ}<\theta<70^{\circ}$ and $5nm>d>0$), and the
phenomenology discussed above is only present in necks with a small
opening angle and large contact diameter.  No jumps in $dV/dI$ are
observed in necks with smaller minimum radius, or necks with larger
opening angle than in the case presented here.  The temperature
dependence of the curves shown in Fig.\ 1 is smooth: the bump
disappears gradually without any sign of peaks in $dV/dI$ as
expected from the decrease of the critical current with
temperature.

Our interpretation gives a rough physical model for the observed
behavior, and agrees only qualitatively with the results.  But it
presents a consistent model and a first approach to understand our
system. We also note that hot spots due to heating
\cite{Klapwijk77} are excluded in our experiment as they should
appear preferentially at zero field, where the bulk is
superconducting and a much poorer heat conductor. We did not
discuss effects associated with the presence of flux tubes as the
flux going through our structure is in any case smaller than the
flux quantum $\Phi_0$.

Recently, other anomalies that appear at zero field in N-S point
contact junctions were explored \cite{Westbrook99}. However, the
position in $V$ of these anomalies changes when decreasing the
smallest contact radius, pointing towards a different origin. Some
nanolithographated structures also present
 anomalies in the current-voltage characteristics whose origin is
debated in Refs.\cite{Strunk,Chien99,Mos94,San91,Park95,Arut99}.
For example, the resistance of some wires does not decrease at
$T_c$ but grows above the normal phase value, and only drops at
lower temperatures.  The differential resistance shows peaks and
anomalies in this regime. The form of these peaks and its
dependence on defects or on the position and nature (normal or
superconducting) of the nanofabricated voltage probes are studied
in detail in Refs. \cite{Strunk,Chien99,Park95,Arut99}. The
presence of phase slip centers and/or a N-S interface also seem to
explain some of the observed behaviors. Also within this context,
the recent theory of Ref.\cite{Devreesse} explains a large critical
field in wedge like structures, but it does not treat the
transition to the resistive state.

Finally, it is also of interest to consider the possibility of
quantum effects, because our necks are indeed potential candidates
to observe quantum of phase slippage (their mean diameter and
length is of the same order of magnitude as the nanowires discussed
in Ref.\cite{Zaikin}). Indeed, recently, the authors of
Ref.\cite{Bezryadin} have measured the destruction of the
superconducting state due to quantum phase slips in wires of
similar diameter and somewhat larger lengths than the connecting
necks presented here. They discussed the possiblity of a
dissipative phase transition (see \cite{Schmid}), depending on the
resistance of the wires. In our case, however, we find anomalies
that appear at high temperatures, indicating that the resistive
centers are activated due to thermal rather than quantum
fluctuations. Future measurements on larger necks and at lower
temperatures might help to further study this point, nevertheless,
we expect that in necks, the presence of the normal (bulk)
electrodes may renormalize the dimensionality of the system and
decrease the importance of (quantum) fluctuations.

In conclusion, we have found a new phenomenology in the transition
to the resistive state of superconducting connecting necks
 of nanoscopic dimensions under magnetic field. We have defined the
conditions for the appearance of different conduction regimes.
Nonequilibrium effects could explain the observed behavior.

Acknowledgments: We acknowledge discussions and help of F.  Guinea, E.
Bascones, A.  Izquierdo, G.  Rubio and N.  Agra\"\i t.  We also acknowledge
finantial support from the TMR program of the EC and the DGICyT under
contracts ERBFMBICT972499 and PB97-0068.

\begin{figure}
\epsfxsize 12cm
\epsfbox{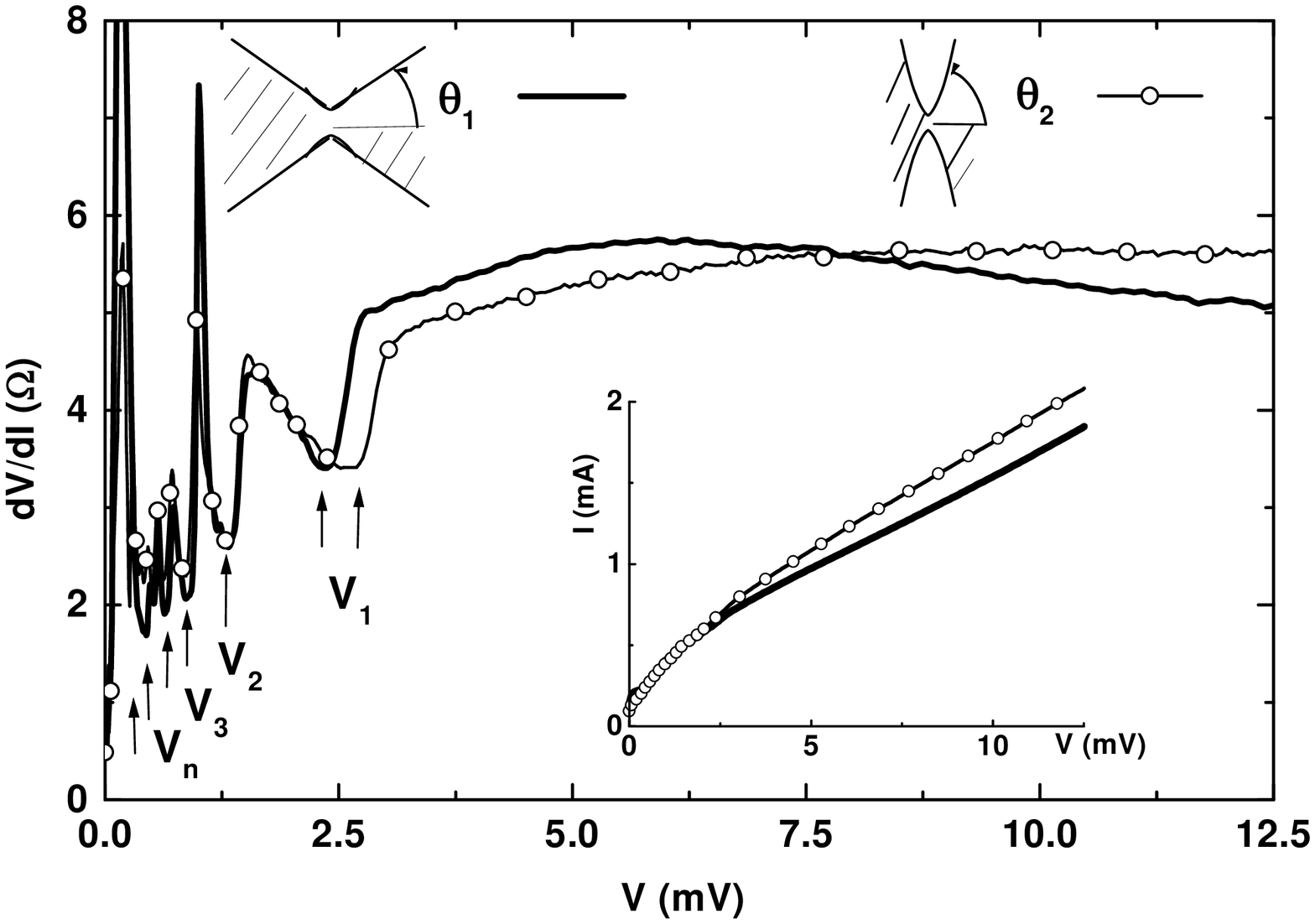}
\caption{The differential resistance $dV/dI(V)$ of a long (line)
and a short (line and points) neck together with the corresponding
I-V curves (inset).   The geometry of the structures, as measured
from the $R_S(z)$ curves, is sketched at the top of the figure.
The arrows show the peaks due to the subharmonic gap structure
\protect\cite{Andreev}.  The peak corresponding to $2\Delta$
appears in the long neck ($\theta_1=35^{\circ }$) at smaller
voltages than in the short neck ($\theta_1=70^{\circ }$) due to
self-heating, which leads to a decrease in $\Delta$.}
\end{figure}

\begin{figure}
\epsfxsize 12cm
\epsfbox{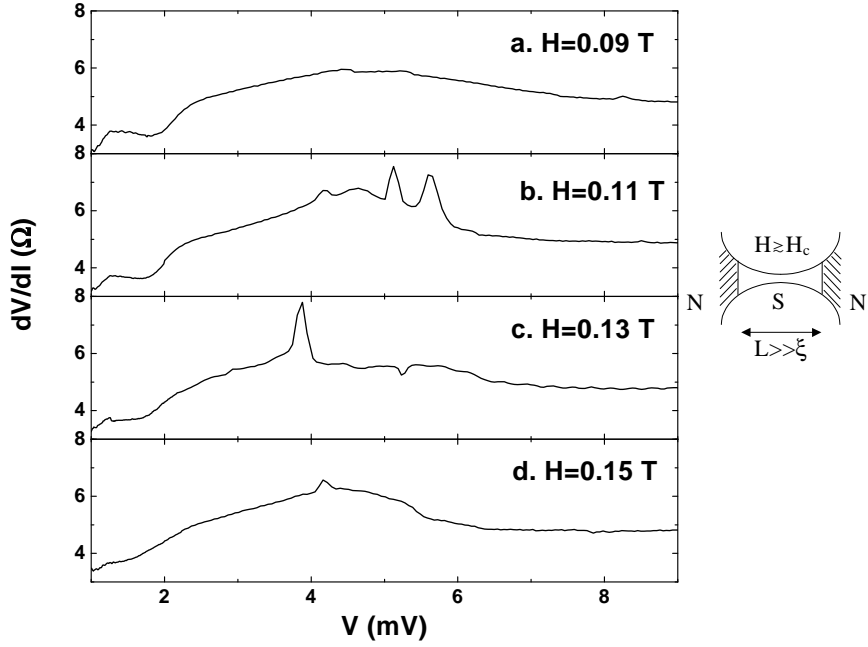}
\caption{ $dV/dI(V)$ for different
magnetic fields ($\theta=35^{\circ}$, see line in Fig.1).  The
sketch on the right is to remark that above $H_c$ ($0.05T$ at 4.2K)
superconductivity is destroyed in the bulk, but not in the neck.
This is discussed in detail in Refs.\protect\cite{Poza98,Sud00}.
For $2H_c<H<3H_c$ we observe high peaks in $dV/dI$ which appear at
on top of the distortion of $dV/dI$ induced by heating.}
\label{fig:Fig2}
\end{figure}

\begin{figure}
\epsfxsize 12cm
\epsfbox{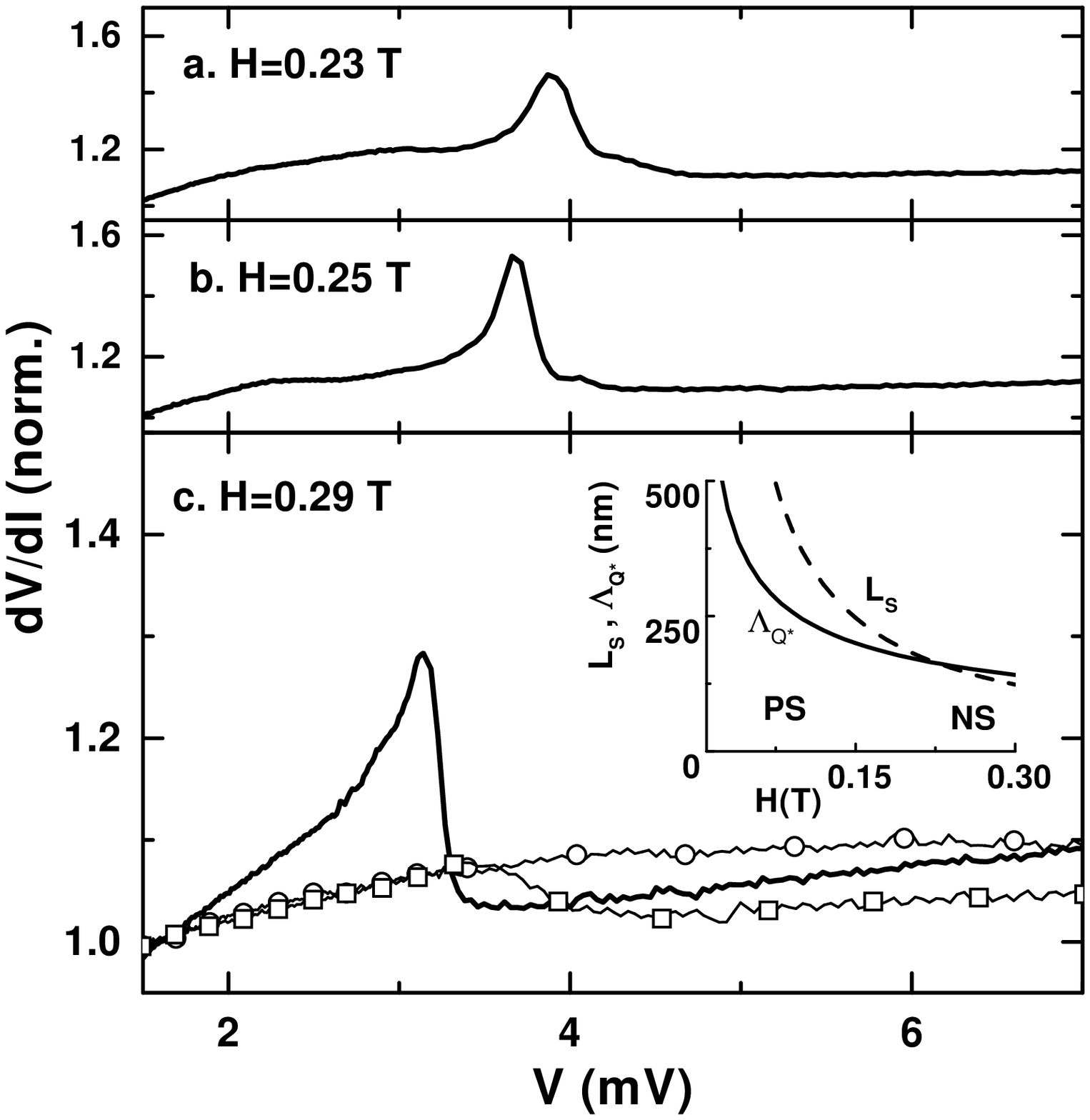}
\caption{
$dV/dI(V)$, normalized to the value in the resistive state as extrapolated
from large voltages.  In c.) we compare the neck showing the anomalous
behavior (line in Fig.1) to a neck with the same $\theta$ but smaller $A$
(ten times smaller, $d=2nm$, squares) and to a neck with two times larger
$\theta$ (open points).  None of these necks show an anomalous behavior.
The inset in c.) sketches the expected decrease of $\Lambda_{Q*}$ and $L_S$
under field.}
\label{fig:Fig3}
\end{figure}


\begin{references}
\bibitem{Checks}
We always check that the $R(z)$ curves have a continuous behavior.
We stretch each neck until it breaks (tunneling regime), and check
subsequently that the topological images of the surface in the
conventional STM mode are the same for different tunneling
resistances.
\bibitem{NotaXi}
Note that he overall length of the neck is limited by the
elongation of the piezotube, 200 nm in our case. This is clearly
several times $\xi$, estimated to be approximately  $20nm$, see
\protect\cite{Sud00}.
\bibitem{comment}
In a neck, we expect an even stronger variation of $\Lambda_{Q*}$
with field, due to the decrease of the length of the
superconducting part, see \cite{Sud00}.
\end{references}
\end{document}